\documentclass[aps,prb,twocolumn,superscriptaddress,showpacs,amsmath,amssymb]{revtex4}
\usepackage{graphicx}
\usepackage{longtable}
\usepackage{bm}

\begin{document}

\title{Pressure and chemical substitution effects in the local atomic structure of BaFe$_2$As$_2$}

\author{E. Granado}
\email{egranado@ifi.unicamp.br}
\affiliation{Instituto de F\'{i}sica ``Gleb Wataghin'', Universidade Estadual de Campinas - UNICAMP, 13083-859, Campinas, SP, Brazil}
\affiliation{Laborat\'{o}rio Nacional de Luz S\'{i}ncrotron, 13083-970, Campinas, SP, Brazil}

\author{L. Mendon\c{c}a-Ferreira}
\affiliation{Instituto de F\'{i}sica e Matem\'{a}tica, Universidade Federal de Pelotas, 96010-900, Pelotas, RS, Brazil}

\author{F. Garcia}
\affiliation{Laborat\'{o}rio Nacional de Luz S\'{i}ncrotron, 13083-970, Campinas, SP, Brazil}

\author{G. de M. Azevedo}
\affiliation{Instituto de F\'{i}sica, Universidade Federal do Rio Grande do Sul (UFRGS), 91501-970, Porto Alegre, RS, Brazil}

\author{G. Fabbris}
\affiliation{Instituto de F\'{i}sica ``Gleb Wataghin'', Universidade Estadual de Campinas - UNICAMP, 13083-859, Campinas, SP, Brazil}
\affiliation{Laborat\'{o}rio Nacional de Luz S\'{i}ncrotron, 13083-970, Campinas, SP, Brazil}

\author{E. M. Bittar}
\affiliation{Instituto de F\'{i}sica ``Gleb Wataghin'', Universidade Estadual de Campinas - UNICAMP, 13083-859, Campinas, SP, Brazil}
\affiliation{Laborat\'{o}rio Nacional de Luz S\'{i}ncrotron, 13083-970, Campinas, SP, Brazil}

\author{C. Adriano}
\affiliation{Instituto de F\'{i}sica ``Gleb Wataghin'', Universidade Estadual de Campinas - UNICAMP, 13083-859, Campinas, SP, Brazil}

\author{T. M. Garitezi}
\affiliation{Instituto de F\'{i}sica ``Gleb Wataghin'', Universidade Estadual de Campinas - UNICAMP, 13083-859, Campinas, SP, Brazil}

\author{P. F. S. Rosa}
\affiliation{Instituto de F\'{i}sica ``Gleb Wataghin'', Universidade Estadual de Campinas - UNICAMP, 13083-859, Campinas, SP, Brazil}

\author{L. F. Bufai\c{c}al}
\affiliation{Instituto de F\'{i}sica, Universidade Federal de Goi\'{a}s, 74001-970, Goi\^{a}nia, GO, Brazil}

\author{M. A. Avila}
\affiliation{Centro de Ci\^{e}ncias Naturais e Humanas, Universidade Federal do ABC, 09210-170, Santo Andr\'{e}, SP, Brazil}

\author{H. Terashita}
\affiliation{Instituto de F\'{i}sica ``Gleb Wataghin'', Universidade Estadual de Campinas - UNICAMP, 13083-859, Campinas, SP, Brazil}
\affiliation{Laborat\'{o}rio Nacional de Luz S\'{i}ncrotron, 13083-970, Campinas, SP, Brazil}

\author{P. G. Pagliuso}
\affiliation{Instituto de F\'{i}sica ``Gleb Wataghin'', Universidade Estadual de Campinas - UNICAMP, 13083-859, Campinas, SP, Brazil}

\begin{abstract}

The effects of K and Co substitutions and quasi-hydrostatic applied pressure ($P<9$ GPa) in the local atomic structure of BaFe$_2$As$_2$, Ba(Fe$_{0.937}$Co$_{0.063}$)$_2$As$_2$ and Ba$_{0.85}$K$_{0.15}$Fe$_2$As$_2$ superconductors were investigated by extended x-ray absorption fine structure (EXAFS) measurements in the As $K$ absorption edge. The As-Fe bond length is found to be slightly reduced ($\lesssim 0.01$ \AA) by both Co and K substitutions, without any observable increment in the corresponding Debye Waller factor. Also, this bond is shown to be compressible ($\kappa = 3.3(3) \times 10^{-3}$ GPa$^{-1}$). The observed contractions of As-Fe bond under pressure and chemical substitutions are likely related with a reduction of the {\it local} Fe magnetic moments, and should be an important tuning parameter in the phase diagrams of the Fe-based superconductors.

\end{abstract}

\pacs{74.70.Xa,61.05.cj,74.62.Fj}

\maketitle

\section{Introduction}

The recent discovery of high-$T_c$ superconductivity in Fe-based pnictides has triggered a large activity in the field.\cite{Paglionereview,Johnstonreview,Canfieldreview} The proximity, and in some cases even the coexistence of superconducting and magnetically ordered states, suggest a similarity of the physics of this system with respect to the cuprates and Ce-based heavy-fermion superconductors. After the initial discovery of electron doped oxypnictides, $R$FeAs(O,F) ($R=$La, rare-earth), with $T_c$'s up to 56 K, \cite{Kamihara,Takahashi,Ren,Wang1} superconductivity was also found in structurally similar materials without oxygen, such as the $A$Fe$_2$As$_2$ ($A=$ Ca, Sr, Ba) with partial chemical substitution,\cite{Rotter1,Rotter2,Sefat1,Sefat2,Canfield2,Canfield3,Jiang,Sharma,Chen,Sasmal,Wu} the LiFeAs family,\cite{Wang2,Pitcher,Tapp} and the iron chalcogenides such as Fe$_{1+x}$Se \cite{Hsu,Mizuguchi} and K$_x$Fe$_2$Se$_2$.\cite{Guo} All these materials share the same structural unit, namely an Fe-As(Se) layer in which each Fe ion is tetrahedrically coordinated with As(Se) ions.

The undoped BaFe$_2$As$_2$ has been regarded as the parent compound of the whole family of Fe pnictides and widely studied, showing a magnetostructural transition from a tetragonal paramagnetic phase into an orthorhombic stripe antiferromagnetic state on cooling below $T_N \sim 140$ K. \cite{Rotter,Huang,Mandrusreview} Either partial K substitution in the Ba site (hole doping),\cite{Rotter1,Rotter2,Chen2} partial substitution of Fe by Co,\cite{Sefat1,Sefat2} Ni,\cite{Canfield2} Rh or Pd,\cite{Canfield3} (electron doping), or even an isovalent P-substitution in the As site\cite{Jiang} or Ru-substitution in the Fe site\cite{Sharma} suppress the magnetic ordering and render this material superconducting.\cite{Canfieldreview} The pure material may also be turned superconducting by application of pressure, with maximum  $T_c \sim 30$ K.\cite{Fukazawa,Alireza2,Colombier,Ishikawa,Matsubayashi,Duncan,Yamakazi} The pressure range in which magnetism is suppressed and superconductivity is observed in BaFe$_2$As$_2$ is highly dependent on the design of the pressure cell\cite{Yamakazi} and on the pressure transmitting medium.\cite{Duncan} In fact, even a small degree of non-hydrostaticity with a correspondingly small uniaxial stress along the $c$ axis strongly suppresses the structural/antiferromagnetic ordering and stabilizes superconductivity at much lower pressures than for perfectly hydrostatic pressures.\cite{Yamakazi,Duncan} 

The evolution of the structural parameters with Co and K substitutions and applied pressure is of great interest, due to the close equivalence between chemical substitution and applied pressure observed in the phase diagram of this family.\cite{Paglionereview,Drotziger} This equivalence has motivated a quest for the relevant parameters that might be tuned both by applied pressure and chemical substitutions.\cite{Kimber,Jorgensen,Drotziger} The most obvious candidate for a relevant structural parameter is arguably the As-Fe bond distance, since it controls directly the chemical pressure on Fe. In addition, it was suggested that the proximity of the As-Fe-As bond angles to the ideal tetrahedral value of 109.47$^{\circ}$ is the main structural parameter tuning superconductivity for BaFe$_2$As$_2$ under pressure and K-substitution.\cite{Kimber} In fact, the Fe pnictides with highest $T_c$'s show nearly ideal As-Fe-As bond angles, suggesting that the potential for high $T_c$'s is enhanced for undistorted FeAs$_4$ tetrahedra.\cite{Johnstonreview,Paglionereview} On the other hand, Drotziger {\it et al.} showed that, in contrast to the application of pressure or doping with K, Ba(Fe$_{1-x}$Co$_x$)$_2$As$_2$ exhibits its $T_c$ maximum for a structure that is far away from that of a regular Fe–As tetrahedron,\cite{Drotziger} indicating that other structural parameters besides  As-Fe-As bond angles may play an important role.

Diffraction studies on BaFe$_2$As$_2$ reported somewhat contradictory effects of an applied pressure on the As-Fe bond distance. First of all, a neutron powder diffraction experiment by Kimber {\it et al.},\cite{Kimber} using a Paris-Edinburgh pressure cell and a mixture of 4:1 methanol:ethanol as pressure-transmitting medium indicated a nearly rigid As-Fe bond under pressure, with a compression  $\lesssim 0.5 $ \% at 150 K for pressures up to 6 GPa. More recently, J\o rgensen and Hansen reported, at room temperature and otherwise similar experimental conditions of Ref. \cite{Kimber}, a larger compressibility of the Fe-As bound ($1.61$ \% at 6.5 GPa).\cite{Jorgensen} Also, Mittal {\it et al.}, by means of x-ray powder diffraction using a diamond anvil cell and He as the pressure transmitting medium, observed a compressible Fe-As bond both at 33 and 300 K ($\sim 4$ \% contraction at 10 GPa).\cite{Mittal} It is also worth mentioning that first principle calculations predict a fairly compressible As-Fe bond.\cite{Xie,Zhang,Kasinathan,Johannes}

Extended X-ray Absorption Fine Structure (EXAFS) measurements may shed light onto the effects of applied pressure and chemical substitutions in the Fe-As bond from a local point of view. It is interesting to note that chemical doping may introduce some degree of disorder in the structure, and therefore the bond distances obtained by diffraction do not necessarily correspond to local values.\cite{Mikkelsen}
In this work, we investigate the As local environment in BaFe$_2$As$_2$ as a function of temperature, Co and K substitutions and quasi-hydrostatic pressure, by means of As $K$ edge EXAFS measurements. Our results indicate that the As-Fe bond distances are slighly reduced by Co and K substitutions, without significant introduction of disorder in the bond direction. A significant local compression of As-Fe bond is also found for the pure compound under application of quasi-hydrostatic pressures. Our results indicate this bond distance may be a relevant structural parameter in destabilizing the itinerant antiferromagnetic ground state and favoring superconductivity in BaFe$_2$As$_2$ and perhaps other similar Fe-based superconductors.

\section{Experimental Details}

\begin{figure}
\includegraphics[width=0.5 \textwidth]{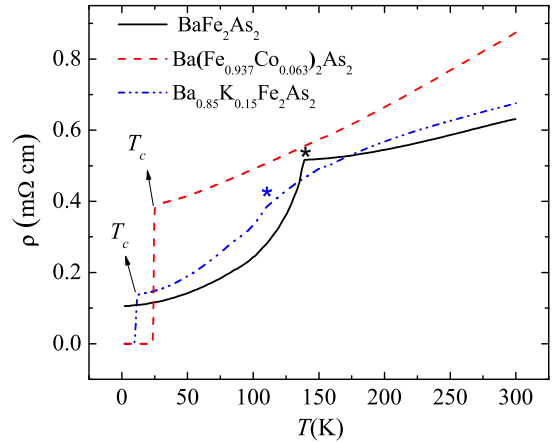}
\caption{\label{resistivity} (Color online) Temperature dependence of the in-plane resistivity of BaFe$_2$As$_2$, Ba(Fe$_{0.937}$Co$_{0.063}$)$_2$As$_2$, and Ba$_{0.85}$K$_{0.15}$Fe$_2$As$_2$. The arrows indicate the onset of superconductivity at $T_c$ and asterisks mark the antiferromagnetic transition temperatures.}
\end{figure}

Single crystals of BaFe$_2$As$_2$ (BFA), Ba(Fe$_{0.937}$Co$_{0.063}$)$_2$As$_2$ (BFCA), and  Ba$_{0.85}$K$_{0.15}$Fe$_2$As$_2$ (BKFA) were synthesized by a flux-growth technique as described in detail elsewhere.\cite{Bittar,Urbano} Thin crystals with linear dimensions in the $ab$ plane between $\sim 0.5$ and $\sim 2$ mm were used. The actual stoichiometry of BFCA was experimentaly determined by the relative Fe and Co $K$ edge jumps in the x-ray absorption spectra.
The compounds were characterized by in plane electrical resistivity ($\rho(T)$) measurements, taken in a commercial platform using the stardard four-probe method (see Fig. \ref{resistivity}). BFA shows a sharp derivative change in $\rho(T)$  at $T_N = 138$ K, marking the simultaneous antiferromagnetic/structural transition, in line with the previously reported $T_N$ for the parent compound.\cite{Rotter,Huang} This indicates the In flux is not significantly incorporated into our BFA sample. For BFCA, no antiferromagnetic transition was observed, and a transition to a superconducting state was found at $T_c = 22$ K. This value actually corresponds to the maximum $T_c$ found in the Co-doped series,\cite{Canfieldreview} and is the expected value for the range of optimal Co-doping. Finally, the BKFA sample shows a superconducting transition at $T_c = 13$ K, and a change of concavity in $\rho(T)$ at $T_N \sim 108$ K, in agreement to the previously reported BKFA phase diagram.\cite{Rotter2,Chen2} Previous nuclear magnetic resonance measurements on a similar sample suggest negligible Sn incorporation in BKFA.\cite{Urbano}

\begin{figure}
\includegraphics[width=0.45 \textwidth]{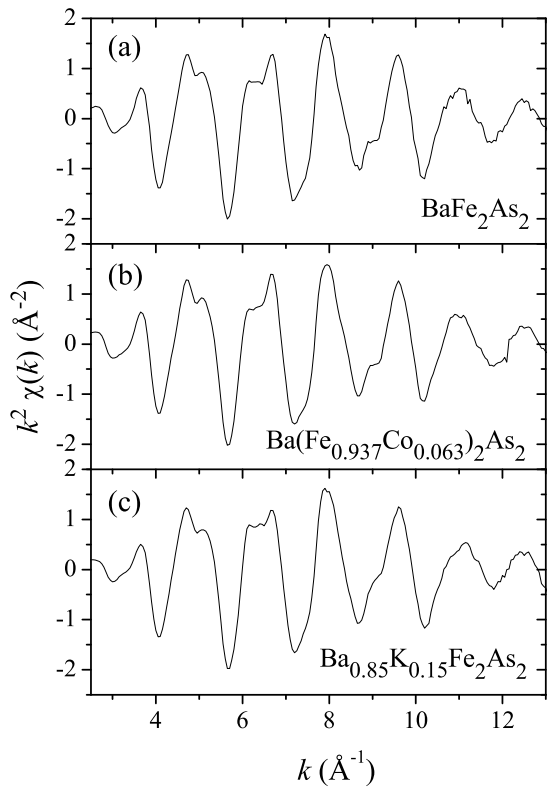}
\caption{\label{spectrak} $k^2$-weighted As $K-$edge extended x-ray absorption fine structure ($k^2 \chi(k)$) spectra of (a) BaFe$_2$As$_2$, (b) Ba(Fe$_{0.937}$Co$_{0.063}$)$_2$As$_2$, and (c) Ba$_{0.85}$K$_{0.15}$Fe$_2$As$_2$, at $T=$298 K.}
\end{figure}

\begin{figure}
\includegraphics[width=0.45 \textwidth]{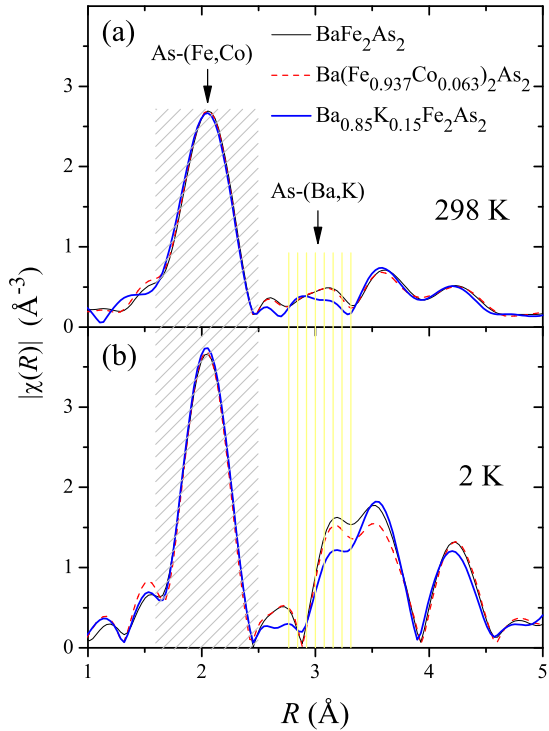}
\caption{\label{spectrar} (Color online) Magnitude of the Fourier transform of As $K-$edge $k^2 \chi(k)$ curves given in Fig. \ref{spectrak}(a-c), ($\chi(R)$), for BaFe$_2$As$_2$, Ba(Fe$_{0.937}$Co$_{0.063}$)$_2$As$_2$, and Ba$_{0.85}$K$_{0.15}$Fe$_2$As$_2$, at (a) $T=$298 K and (b) $T=2$ K. The intervals corresponding approximately to the first [As-(Fe,Co)] and second [As-(Ba,K)] As coordination shells are indicated as dashed areas.}
\end{figure}

\begin{figure}
\includegraphics[width=0.45 \textwidth]{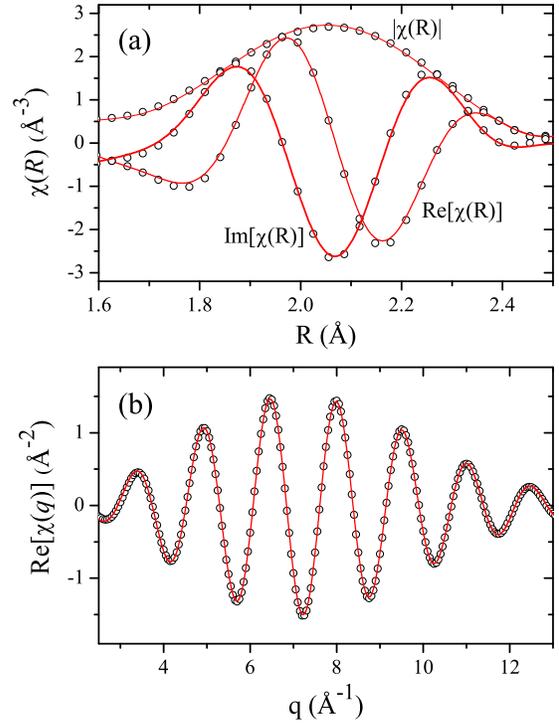}
\caption{\label{spectrafit} (Color online) (a) Envelope, real and imaginary components of $\chi (R)$ for BaFe$_2$As$_2$, at $T=$298 K, in the first shell interval. (b) Real component of the backward Fourier transform of $\chi (R)$ ($1.6$ \AA\ $<R<2.5$ \AA\ ), $\chi (q)$. Circles: experimental data; solid lines: calculated.}
\end{figure}

\begingroup
\begin{table*}
\caption{\label{table1} Refined As-$M$ ($M=$Fe,Co) distances and Debye-Waller factors obtained from the fits of x-ray absorption fine structure data at the As $K-$edge at ambient pressure. Errors given in parentheses are statistical only, and are defined as the standard deviation of the results obtained from repeated measurements under identical conditions.}
\begin{ruledtabular}
\begin{tabular}{c c c c} 
& $T=2$ K & $T=30$ K & $T=298$K  \\
\hline
BaFe$_2$As$_2$ & & &  \\
$d$(As-Fe) (\AA) & 2.3915(12) & 2.3914(7)& 2.3985(14) \\
 $\sigma^2$ (\AA\ $^2$) & 0.00266(12) & 0.00250 (7) & 0.00465(11) \\
\hline
Ba[Fe$_{0.937}$Co$_{0.063}$]$_2$As$_2$ & & & \\
$d$[As-(Fe,Co)] (\AA) & 2.3833(12) & 2.3838(9) & 2.3951(12) \\
$\sigma^2$ (\AA\ $^2$) & 0.00262(12) & 0.00268(9) & 0.00466(9) \\
\hline
Ba$_{0.85}$K$_{0.15}$Fe$_2$As$_2$ & & & \\
$d$(As-Fe) (\AA) & 2.3865(15)& 2.3900(12) & 2.3955(9) \\
$\sigma^2$ (\AA\ $^2$) & 0.00242(15)& 0.00248(12)& 0.00466(7) \\

\end{tabular}
\end{ruledtabular}
\end{table*}
\endgroup

\begin{figure}
\includegraphics[width=0.45 \textwidth]{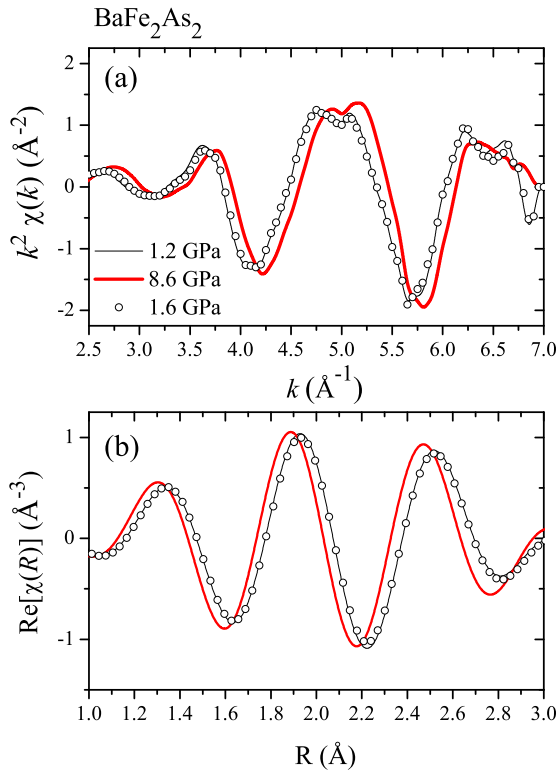}
\caption{\label{spectrap} (Color online) (a) $k^2$-weighted As $K-$edge extended x-ray absorption fine structure ($k^2 \chi(k)$) spectra of BaFe$_2$As$_2$ at 298 K, with the sample loaded into the diamond anvil cell, with $P=1.2$ GPa (thin line), $P=8.6$ GPa (thick line), and $P=1.6$ GPa after release from 8.6 GPa  (circles). (b) Real component of the Fourier transform of $k^2 \chi(k)$ curves given in (a).}
\end{figure}

\begin{figure}
\includegraphics[width=0.5 \textwidth]{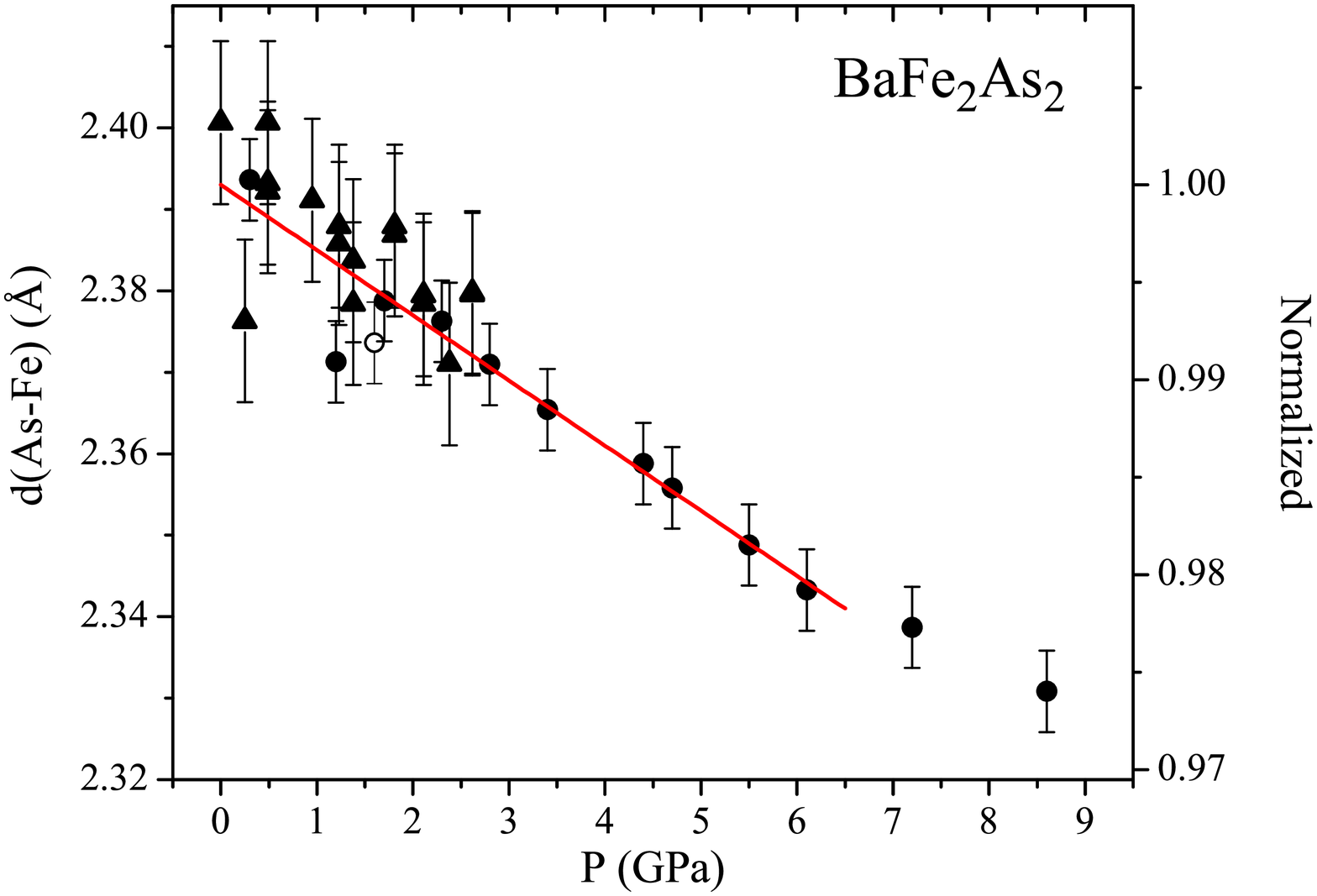}
\caption{\label{pressureresults} (Color online) Pressure-dependence of the As-Fe bond distance at 298 K for BaFe$_2$As$_2$, obtained from the fits to the As $K$ edge x-ray absorption fine structure (see Fig. \ref{spectrap}). Circles and triangles represent data taken in different runs. Data were taken for increasing pressures, except for the open circle at 1.6 GPa, which was obtained after releasing the cell from 8.6 GPa. Errorbars are statistical only, and were estimated from the standard deviation of the results obtained from repeated measurements under identical conditions. The solid line is a linear fit to the data below 6.1 GPa, representing a Fe-As bond compressibility of $\kappa = 3.3(3) \times 10^{-3}$ GPa$^{-1}$.}
\end{figure}

Preliminary x-ray diffraction measurements revealed that manual sample grinding lead to significant Bragg peak broadening, suggesting this procedure may  introduce significant strain into the crystal structure, with unpredictable effects on the local atomic and electronic structures. Therefore, only single crystals were used in this EXAFS study, including the high-pressure measurements. The natural thicknesses of selected crystals along the $c$ direction were found to be appropriate to perform high-quality EXAFS measurements in transmission mode at the As $K$ edge ($E=11865$ eV). On the other hand, the crystals were too thick for transmission EXAFS measurements at the Fe and Co $K$ edges ($E=7116$ and 7713 eV, respectively), and therefore the methodology employed in this work for the As $K$ edge was not directly extended to the transition metal $K$ edges.

Temperature-dependent EXAFS measurements at the As $K$ edge were performed at ambient pressure in the XAFS-2 beamline at the Brazilian Synchrotron Laboratory (LNLS) in transmission mode.  The samples were placed into a closed-cycle He cryostat in series with a open-cycle Joule-Thompson He circuit, yielding a base temperature of $\sim 1.7$ K. The samples were mounted into a Be dome filled with He gas to account for proper heat exchange and temperature homogeneity. Temperature stability was better than 0.01 K and precision was better than 0.1 K. An Au foil was used for energy calibration.

Pressure-dependent EXAFS measurements at ambient temperature were performed with dispersive optics in the DXAS beamline of LNLS,\cite{DXAS} using a diamond anvil cell. The samples were loaded into a hole of an iconel gasket with diameter of $\sim 200$ $\mu$m. Ruby spheres of $\sim 40$ $\mu$m-diameter were also loaded into the cell. An admixture of methanol, ethanol, and water in the proportion 16:3:1 was used as pressure-transmitting media.\cite{Fujishiro} The applied pressures were optically measured offline by the position of the ruby fluorescence doublet. These lines present no observable broadening with increasing pressures, demonstrating quasi-hydrostatic pressures over the whole investigated pressure interval.

The raw EXAFS measurements were pre-processed to obtain $\chi(k)$ by conventional methods using the software ATHENA.\cite{Ravel} Theoretical signals for atomic shells were derived using the FEFF8 code,\cite{Ankudinov} taking the reported tetragonal crystal structure of BFA at 175 K (space group $I4/mmm$)\cite{Huang} as the initial model. The fits were performed in the real space, by optimizing the Fourier transform of $k^2 \chi(k)$, using the IFEFFIT program\cite{Newville} running under the ARTEMIS graphical platform.\cite{Ravel} For the temperature-dependent measurements at the As $K$ edge, fits of the As-Fe distance were confined to the $k$ range of $3 <k<13$ \AA\ $^{-1}$
and to $R$ range of $1<R<3$ \AA, while for the pressure-dependent data the useful $k$ range was significantly shorter, $2.5<k<6.5$ \AA\
$^{-1}$, and the employed $R$ range was $1.2 <R <4.2$ \AA\ (see below).


\section{Results and analysis}

According to the crystal structure of BaFe$_2$As$_2$,\cite{Huang} each As ion is surrounded by four Fe (or Co) nearest-neighbors, defining the first As coordination shell with $d_{As-Fe}$ $\sim 2.4$ \AA. In both the tetragonal (space group $I4/mmm$) and orthorhombic ($Fmmm$) phases, a single As-Fe bond distance is defined. The As second nearest neighbors are the Ba (or K) ions, with a significantly larger $d_{As-Ba} \sim 3.4$ \AA. The As-As distances (third shell) are $d_{As-As} \sim 3.8-4.0$ \AA. This fact allows the EXAFS signal arising from the As-(Fe,Co) first shell to be unambiguously isolated and analysed, making EXAFS at the As $K$ edge an ideally suited technique to study the As-(Fe,Co) bond in these materials.

Figures \ref{spectrak}(a-c) show the As $K$ edge $k^2$-weighted raw EXAFS data ($k^2 \chi(k)$) at 298 K for BFA, BFCA, and BKFA, respectively. The magnitude of the Fourier transform of $k^2 \chi(k)$ into the real space ($\chi(R)$) is given in Fig. \ref{spectrar} for all studied compounds at room temperature and 2 K. These data show an  amplitude peak in $\chi(R)$ centered at the non-phase-corrected radial distance $R \sim 2.05$ \AA, which is readily associated with the As-(Fe,Co) first shell. The region $2.5 \lesssim R \lesssim 3.3$ \AA, associated to the As-(Ba,K) coordination shell, is substantially altered for the BKFA sample with respect to BFA and BFCA, due to the largely different scattering amplitudes of Ba and K. Finally, all studied compounds show similar features in $\chi(R)$ for $R>3.3$ \AA.

The envelope, real and imaginary components of $\chi (R)$ for BFA at 298 K are shown in Fig. \ref{spectrafit}(a). A fit of these data to a simple first-shell model in the region of interest is illustrated in Figs. \ref{spectrafit}(a). Figure \ref{spectrafit}(b) shows the nearly perfect fit to the backward Fourier transform data of $\chi (R)$ in the first shell interval, $1.6$ \AA\ $<R<2.5$ \AA\  ($\chi (q)$). We should mention that the fit quality shown in Figs. \ref{spectrafit}(a,b) is similar for other compounds and temperatures.

Table \ref{table1} shows the refined As-(Fe,Co) bond distances and Debye-Waller factors ($\sigma ^2$) for BFA, BFCA, and BKFA, at 2, 30, and 298 K. For BFA, the As-Fe bond distance reported here at 2 K is consistent with that obtained by neutron powder diffraction at 5 K,\cite{Huang} within experimental errors. Both Co- and K-substitution lead to a slight compression of the As-(Fe,Co) bond ($\lesssim 0.01$ \AA) with respect to the pure compound. Also, no systematic increase of the $\sigma ^2$ for this bond may be noticed for the substituted compounds. As expected, the As-(Fe,Co) bond distances decrease slightly on cooling from room temperature down to 2 K, for all investigated samples. Also, $\sigma ^2$ for this bond is sensibly reduced on cooling for all studied compounds and consistent with previously reported values,\cite{Cheng} reaching values close to $\sigma^2 = 0.0025$ \AA\ $^{2}$ at 2 K for all studied compounds. No variations of As-(Fe,Co) distance or $\sigma ^2$ between the superconducting and normal states at 2 and 30 K, respectively, were observed for BFCA within experimental errors. For BKFA, a possible As-(Fe,Co) bond elongation of 0.0035(19) \AA\ $^{2}$ takes place between 2 K and 30 K. Since this possible elongation is within two standard deviations and was observed for only one sample, we do not ascribe statistical significance to it in this work.

Figure \ref{spectrap}(a) shows $k^2 \chi(k)$ for pure BFA taken in dispersive geometry at the As $K$ edge with the sample inside the diamond anvil cell, at room temperature and selected pressures. The setup used in this high-pressure experiment restricts the accessible $k$-range. Within this range, the spectrum seen in Fig. \ref{spectrap}(a) is consistent with the one shown in Fig. \ref{spectrak}(a). The real part of the Fourier transform of  $k^2 \chi(k)$ into the real space is given in Fig. \ref{spectrap}(b). Due to the limited $k$-range, the resolution in $R$-space is degraded with respect to the data shown in Fig. \ref{spectrar}. However, since the As first shell is completely separated from the second shell, an analysis of the local As-Fe bond with pressure is still possible. Clearly, the application of pressure leads to a phase change of $k^2 \chi(k)$, corresponding to a reduction of the Fe-As bond distance. These trends are reversible, at least for pressures up to 8.6 GPa, which is the maximum value reached in this study. Figure \ref{pressureresults} shows the pressure-dependence of the refined As-Fe bond distance. From a linear fit to these data below 6.1 GPa, a compressibility $\kappa = 3.3(3) \times 10^{-3}$ GPa$^{-1}$ is obtained for this bond.  

For the As second coordination shell [As-(Ba,K) bond], the analysis becomes much more ambiguous, due a significant overlap with the third shell (As-As bond) and with a multiple scattering path (As-Fe-Fe-As). In addition, the two independent As-Ba distances expected for the orthorhombic $Fmmm$ phase and the possibly different local As-K and As-Ba distances for BKFA bring further complexity to the analysis of the second coordination As shell, which is not relevant to our conclusions and beyond the scope of the present work. 

\section{Discussion}

The As-(Fe,Co) bond distances and Debye-Waller factors ($\sigma^2$) displayed in Table \ref{table1} bring insight into the effects of chemical electron (Co) and hole (K) doping on the local atomic structure of BaFe$_2$As$_2$. First of all, the nearly identical $\sigma^2$ for all studied samples reveal no observable  disorder in this bond length brought by either K, and, most remarkably, Co substitution. The absence of a significant structural disorder in the (Fe,Co)As layers is possibly a key factor to favor superconductivity in these materials under doping.

An additional relevant information extracted from Table \ref{table1} is the tendency for slighly shorter As-(Fe,Co) bonds for both hole and electron doped samples with respect to pure BaFe$_2$As$_2$. In addition, the pressure-dependence of the As-Fe distance (Fig. \ref{pressureresults}) shows a significant compressibility. Therefore, Co and K substitutions, as well as application of pressure, produce at least one common qualitative structural trend, i.e., a shortening of the As-(Fe,Co) bond. For the specific case of BFCA, the shortening of the As-(Fe,Co) bond of $\sim 0.01$ \AA\ at low temperatures with respect to pure BFA might be, in principle, attributed to smaller As-Co bond lengths with respect to As-Fe. However, the small Co concentration for this sample and the negligible bond disorder brought by Co substitution mentioned above dismiss this hypothesis, and we conclude that a significant local contraction of the As-Fe bond is induced by Co substitution, such as in the case of K substitution and applied pressure. 

Ab-initio band structure calculations based on the density functional theory described the predicted structural parameters and local Fe magnetic moment in BaFe$_2$As$_2$ as a function of applied pressure\cite{Xie,Zhang, Kasinathan,Johannes} and chemical substitution/charge doping.\cite{Kasinathan,Johannes} Particularly, it was shown that the As-Fe bond length is a gauge of the magnitude of the {\it local} magnetic moment in the Fe site.\cite{Johannes} Thus, the contraction of the As-Fe bond length reported here indicates a reduction of the Fe moments upon application of pressure and Co/K substitutions. This trend is in qualitative agreement with the first principles calculations.\cite{Kasinathan,Johannes} We should mention that the slightly larger compressibility observed for the As-Fe bond with respect to the calculated one ($\kappa_{calc} = 2.5 \times 10^{-3}$ GPa$^{-1}$)\cite{Johannes} is likely related with an overestimation of the calculated static magnetic moments by density functional theory.\cite{Johannes}

It is concluded from the above considerations that a decrease of the As-Fe bond length, induced either by pressure or selected chemical substitutions, may be an important feature that accompanies the destabilization of the magnetic order and emergence of superconductivity in BaFe$_2$As$_2$. We should mention that other structural parameters, such as the As-Fe-As bond angles and Fe-Fe distances, may also influence the equilibrium between the superconducting and magnetically ordered ground states, through the control of the interatomic exchange interactions, albeit the As-Fe distance is more directly related with the local Fe moments.

\section{Conclusions}

In summary, our EXAFS results show that both electron-doping, introduced by Co substitution in the Fe site, and hole-doping, associated with K substitution in the Ba site, lead to a small reduction of the As first shell bond length at low temperatures with respect to the pure compound, without introducing any observable additional disorder in this shell. In addition, it is demonstrated that application of quasi-hydrostatic pressure in the pure compound is responsible for a significant compression of the As-Fe local bond. These observations, interpreted in the light of the direct relationship between the As-Fe distance and the local Fe magnetic moments,\cite{Johannes} 
provide insights into the similarity between chemical doping and applied pressures in destabilizing the magnetic order and favoring the superconducting ground state in BaFe$_2$As$_2$.

\section{Acknowledgements}

This work was supported by Fapesp and CNPq, Brazil. LNLS is acknowledged for concession of beamtime.

\end{document}